\documentclass[reprint,superscriptaddress,amsmath,amssymb,aps,pra]{revtex4-2}

\usepackage{graphicx}
\usepackage{dcolumn}
\usepackage{bm}
\usepackage{blindtext}
\usepackage{xcolor}
\usepackage{lineno}
\usepackage{tabulary,booktabs}
\usepackage{amsmath}
\usepackage{wrapfig}
\usepackage{mathtools}
\usepackage{comment}
\usepackage{chngcntr}

\begin{document}

\preprint{APS/123-QED}

\title{Local sensing of gas microflows using a levitated nanoparticle}

\author{Maxime Debiossac}
\thanks{Equal contribution.}
\email{maxime.debiossac@dlr.de}
\affiliation{Vienna Center for Quantum Science and Technology (VCQ), Faculty of Physics, University of Vienna, A-1090 Vienna, Austria}
\affiliation{German Aerospace Center (DLR), Institute of Quantum Technologies, Wilhelm-Runge-Strasse 10, 89081 Ulm, Germany}

\author{David Grass}
\thanks{Equal contribution.}
\affiliation{Vienna Center for Quantum Science and Technology (VCQ), Faculty of Physics, University of Vienna, A-1090 Vienna, Austria}
\affiliation{Department of Chemistry, Duke University, Durham, NC 27708, USA}

\author{Daniel R. Ladiges}
\affiliation{Center for Computational Sciences and Engineering, Lawrence Berkeley National Laboratory, Berkeley, California 94720, USA}
\affiliation{ARC Centre of Excellence in Exciton Science, School of Mathematics and Statistics, The University of Melbourne, Victoria 3010, Australia}

\author{John E. Sader}
\affiliation{Lynn Booth and Kent Kresa Department of Aerospace \& Department of Applied Physics and Materials Science,
California Institute of Technology, Pasadena, CA 91125, USA}

\author{Markus Aspelmeyer}
\affiliation{Vienna Center for Quantum Science and Technology (VCQ), Faculty of Physics, University of Vienna, A-1090 Vienna, Austria}

\author{Nikolai Kiesel}
\affiliation{Vienna Center for Quantum Science and Technology (VCQ), Faculty of Physics, University of Vienna, A-1090 Vienna, Austria}

\date{\today}

\begin{abstract}
Local measurements of gas flow in microscale geometries remain challenging due to the difficulty of micromechanical sensors implantation. Here, we use a levitated nanoparticle as a minimally invasive probe to map, with high spatial resolution, both the local pressure and velocity gas flow for a canonical example - a micropipe subjected to a pressure gradient. This technique enables characterization of nonequilibrium gas transport across several flow regimes, from slip flow to free-molecular flow. We directly observe the emergence of a nonlinear pressure distribution at large pressure ratios and obtain agreement with kinetic-theory predictions and Direct Simulation Monte Carlo calculations. These results introduce levitated nanoparticles as powerful and reconfigurable probes for microchannel gas sensing, with applications in microtechnology, micro-chemistry, and bioengineering.
\end{abstract}


\maketitle

Gas flows in microchannels are ubiquitous in science and technologies. They represent the building blocks of microsystems with applications in various areas such as mechanical and electrical engineering, chemistry, medicine, biotechnology, optics and aerospace~\cite{fujita2002microactuators}. Understanding gas flows at the microscale is for example essential to designing micro-mechanical systems (MEMS) such as micro motors~\cite{rapp1994liga}, micro actuators~\cite{thielicke2000microactuators} or micro sensors~\cite{wang2009mems}. In chemistry, gas microflows are key for micro heat exchanger in chemical reactions~\cite{ohadi2013next} and for performing heterogenous catalysis in gas micro reactors~\cite{tonkovich1999microchannel}. In biology and medicine, gas micro chromatograph~\cite{ohira2008micro} and micro mass spectrometers~\cite{badman2000miniature}  for environmental and medical gas analysis are currently an active field of research. 

Despite the tremendous progress in gas microflows, it remains challenging, even for simple geometry, to probe gas velocity and pressure flow locally inside microchannels due to the difficulty of micro sensor implantation. In this work, we consider the paradigmatic example of gas flow of air molecules through a micropipe connecting two reservoirs at different pressures.  
Gas flow in micropipes exhibits nonequilibrium effects such as rarefaction and gas-surface interaction that differ significantly from their macroscopic counterparts~\cite{beskok1996rarefaction}. In addition, depending on the pressure gradient along the micropipe, several flow regimes may coexist~\cite{yang2009rarefied}, ranging from  (continuum) viscous flow governed by the Navier-Stokes equation to rarefied flow described by kinetic theory. The Boltzmann equation~\cite{boltzmann1970weitere}, one of the most important equations of nonequilibrium statistical mechanics, is applicable to all flow regimes~\cite{lei2016physics}. However, due to its nonlinear nature, 
exact analytical solutions of the Boltzmann equation are scarce in general, even for a priori simple geometries such as circular microchannels. Previous works have derived approximate models aiming at covering the entire range of gas flow~\cite{mccoy1974transport,beskok1999report,arkilic2002gaseous,weng1999gaseous}. They were, however, not confronted by experiments for large pressure gradients, a situation only considered in Ref.~\cite{yang2009rarefied}.

\begin{figure*}[t]
    \centering
    \includegraphics[width=0.8\paperwidth]{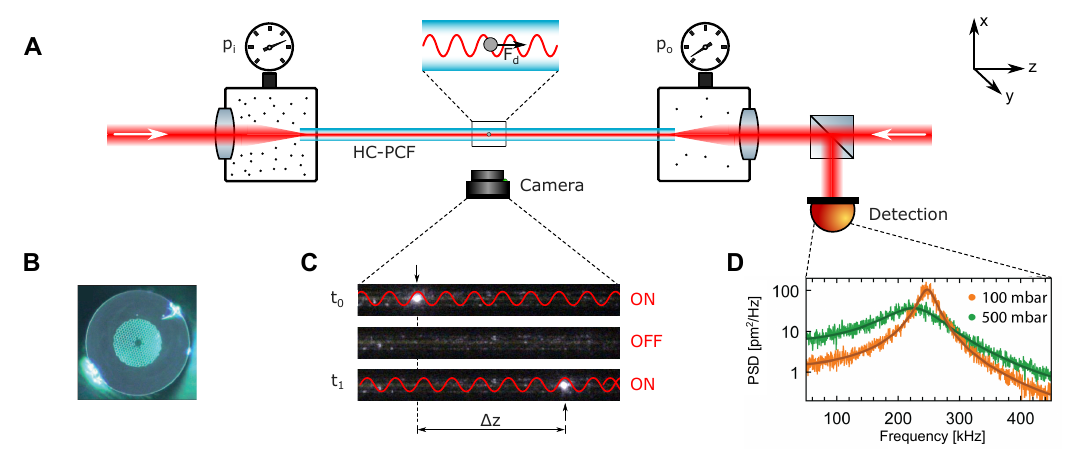}
    \caption{\textbf{Experimental setup. (A)} A $L = 18.5\pm 0.5$ cm long hollow core photonic crystal fiber (HC-PCF) with a hollow-core radius $R = 5\pm 0.5~\mu$m connects two vacuum chambers, filled with air at pressure $p$, and maintained at pressure $p_i$ (inlet) and $p_o$ (outlet). A dielectric particle is trapped inside the HC-PCF using two counterpropagating light beams. \textbf{(B)} Dark field microscope image of a cross section of the HC-PCF used in this setup. \textbf{(C)} Camera images of the levitated particle inside the HC-PCF during a flow velociy measurement. The particle is trapped in an intensity maximum of the standing wave at time $t_0$, dragged by air flow along a distance $\Delta z$ when the trap is switched off and being recaptured at time $t_1$ after switching the light back on. The local gas velocity is estimated as $v(z,r)=\Delta z/(t_1-t_0)$. \textbf{(D)} Power spectral density (PSD) of the particle center of mass motion recorded using a photodiode. The local pressure surrounding the particle is inferred from the PSD of the particle motion. Orange (green) data in the inset corresponds to a particle trapped at a pressure of $p_{i,o}=$ 100 mbar (500 mbar).}
    \label{fig:setup}
\end{figure*}

To date, most experimental studies of gas flow in micropipes are restricted to measurements of volumetric flow rate and pressure at the inlet and outlet~\cite{lei2016physics,ewart2006mass}. This is mainly due to intrusiveness issues, i.e. the difficulty of microsensor implantation inside micro-fluidic devices~\cite{brandner2019situ}. The amount of experimental data on gas microflows is therefore very limited. Few works have measured the local pressure profiles along a microchannel, using either integrated micro-machined pressure sensors~\cite{liu1995mems,zohar2002subsonic,hsieh2004gas} or the quenching property of luminescent molecules by surrounding oxygen molecules~\cite{huang2007microchannel}. Nonlinear pressure profiles due to gas compressibility  were  reported~\cite{pong1994non,huang2007microchannel,gao2018experimental}, although only for small pressure gradients between the inlet and outlet of the microtube, supporting mainly only one flow regime. 

In this letter, we employ levitated particles to measure both the local pressure and flow velocity of gas molecules along a microchannel---a Hollow Core Photonic Crystal Fiber (HC-PCF)---subjected to a large pressure gradient. Our experimental results are supported by Direct Simulation Monte Carlo data. We also provide analytical solutions for the pressure and flow velocity distributions along the microchannel which cover all flow regimes.

HC-PCF offer an attractive interface for confinement and deployment of laser light in a shielded environment and over long ($\sim$ km) distances~\cite{knight2002new,russell2003photonic}. Due to their flexibility, compact structure and fast response, HC-PCF have been used in the field of ultrafast lasers~\cite{russell2014hollow,safaei2020high,couny2007generation}, gas photonics~\cite{benabid2002stimulated,masum2019numerical} and quantum optics~\cite{tyumenev2022tunable,epple2014rydberg,takekoshi2007optical,christensen2008trapping,epple2014rydberg}. More recently, HC-PCF have been exploited to guide and trap particles with micrometer positional accuracy using optical~\cite{benabid2002particle,schmidt2011metrology} and opto-thermal~\cite{schmidt2012reconfigurable} forces.  This has opened new avenues in particle-based sensing with applications such as electrometry~\cite{bykov2015flying}, thermometry~\cite{bykov2015flying} and magnetometry~\cite{chakraborty2024optomagnetic}. Here, we extend the sensing capabilities of HC-PCF to pressure and velocity flow sensing using a levitated particle.

\section*{Distributed gas flow sensor}

The experimental arrangement is depicted in Fig.~\ref{fig:setup}. An HC-PCF with a core radius $R=5\pm0.5~\mu$m and length $L=18.5\pm0.5$ cm connects two vacuum chambers, in which the pressure (air) can be controlled independently. The two vacuum chambers and the HC-PCF are kept at room temperature. A silica particle is trapped at the intensity maximum of a standing wave, formed by two counterpropagating laser beams ($\lambda=1064$ nm) coupled into the HC-PCF. The particle can be positioned precisely along the HC-PCF by frequency detuning one of the laser beams, thereby forming an optical conveyor belt~\cite{Grass2016}. Such optical control allows us to  transport and position the particle inside the HC-PCF. We can track the particle motion by observing the interference between the scattered and trapping light fields. The motional power spectral density allows us to probe locally the pressure $p(z)$ and gas velocity distribution $v(z,r)$ of air molecules along the fiber submitted to a pressure gradient $\Pi=p_i/p_o$. Here $p_o$ ($p_i$) is the outlet (inlet) pressure. Depending on this pressure gradient, several categories of gas flow can coexist along the fiber. They are classified into four regimes according to the Knudsen number $\text{Kn}=\lambda/R$ that relates the molecular mean free path of gas $\lambda=\mu_0/p\sqrt{\pi R_sT/2}$ to a characteristic dimension
of the microchannel (the radius $R$). Here, $\mu_0=1.81\times 10^{-5}$ Pa.s is the air viscosity, $T=300$ K is the temperature and $R_s$ is the specific gas constant. We distinguish the following flow regimes~\cite{schaaf1961flow}\\

\noindent
\begin{tabular}{ll}
\hline
Continuum flow &  $\text{Kn}\lesssim10^{-3}$   \\
Slip flow &  $10^{-3}\lesssim\text{Kn}\lesssim10^{-1}$ \\
Transition flow & $10^{-1}\lesssim\text{Kn}\lesssim10$ \\
Free molecular flow & $\text{Kn}\gtrsim10$ \\
\hline
\end{tabular}

\noindent
We present our results for $p_i=(300,500,700)$ mbar and $p_o=10$ mbar, covering the slip flow and transition flow regimes ($0.02<\text{Kn}<1.4$). The variation between the transition flow and the free molecular flow regimes ($0.1<\text{Kn}<67$) is detailed in Fig. S5 in Supplementary Materials. 

\section*{Local pressure gas flow distribution}

Levitated particles offer exquisite pressure sensing capabilities through their interaction with the background gas~\cite{gonzalez2021levitodynamics}. The particle motion inside the HC-PCF is damped by the surrounding gas molecules 
with a damping rate $\gamma(p)=6\pi a K\mu_0 f(p)/m$ where $a$ is the particle radius, $m$ its mass and $f(p)$ accounts for the degree of rarefaction. Here, $K\approx 1+2.105 a/R$ is a correction factor that accounts for the effect of the fiber walls on the particle~\cite{brenner1961slow}. The rarefaction function $f(p)$ is related to the local pressure $f(p)=0.619\left(1+c_K\right)/(0.619+\text{Kn}_c)$~\cite{beresnev1990motion},
where $\text{Kn}_c=\lambda/L_c$ is the particle Knudsen number with $L_c$ its characteristic length and $c_K= 0.31\text{Kn}_c/(0.785+1.152\text{Kn}_c+\text{Kn}_c^2$). The damping rate $\gamma(p)$ can be extracted from the power spectral density of the particle motion $S(\omega)\propto \gamma/\left[(\Omega^2-\omega^2)^2+\gamma^2\omega^2\right]$ with $\Omega$ the particle frequency (Fig.~\ref{fig:setup}D). The gas friction law above can then be inverted to provide with the local pressure distribution $p(z)$ along the HC-PCF (see Fig. S7 in the Supplementary Materials). 

\begin{figure}[t]
    \centering
    \includegraphics[width=0.9\columnwidth]{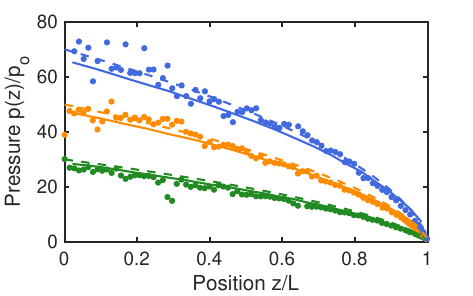}
    \caption{\textbf{Local flow pressure distribution.} Pressure profiles $p(z)$ measured along the HC-PCF with levitated particle ($a=416$ nm). In all cases the outlet pressure is $p_o=10$ mbar and the inlet pressure is $p_i=$ 300 mbar (green), 500 mbar (orange) and 700 mbar (blue). Full circles corresponds to experimental data points extracted from the power spectral density of the particle motion. Solid lines are the DSMC simulation data and the dashed lines are the analytical formula from Eq.~\ref{eq:pressure}.}
    \label{fig:pressure}
\end{figure}

In Fig.~\ref{fig:setup}D, we show the power spectral density of a particle inside the fiber for $p_{i,o}=100$ mbar (orange) and 500 mbar (green). Such measurements, without a pressure gradient along the fiber, provide a simple calibration of the pressure sensor (see Fig. S7). To measure the local pressure $p(z)$ in the presence of a pressure gradient, we move the particle using the optical conveyor belt at a position $z$ along the fiber and record a power spectrum density. The pressure distribution $p(z)$ is shown in Fig.~\ref{fig:pressure} for various pressure ratios $\Pi=30$ (green), 50 (orange), and 70 (blue) with an outlet pressure fixed at $p_o=10$ mbar for all measurements. The pressure distributions inside the HC-PCF exhibit a nonlinear behavior stemming from the large pressure ratio.  This nonlinearity scales with the pressure ratio, highlighting the effect of flow compressibility. To support our experimental results, we have performed Direct Monte Carlo simulations (DSMC, see Supplementary Materials). For pressures bellow 300 mbar, we observe a good agreement between the experimental data and the DSMC results for all pressure ratios. Notice in particular that for $\Pi=30$, the experimental data correctly capture the pressure drop at the inlet of the fiber. This pressure drop, referred as the entrance effect~\cite{huang2007microchannel,jang2004pressure}, decreases with the Knudsen number Kn and becomes more significant for large pressure ratio $\Pi$~\cite{huang2007microchannel}. The entrance effect is, however, less visible for $\Pi=50$ and 70 due to the decreased motional sensitivity of the particle at higher pressures, that is, close to the fiber inlet. We also compare our results to the pressure distribution obtained for long microtubes in the 
slip flow boundary condition (see Supplementary Materials):
\begin{equation}
    \tilde{p}(\tilde{z})=-\kappa_o+\sqrt{(\Pi+\kappa_o)^2+(1-\Pi)(1+\Pi+2\kappa_o)\tilde{z}}.
    \label{eq:pressure}
\end{equation}
Here, $\tilde{p}=p/p_o$, $\tilde{z}=z/L$, $\Pi=p_i/p_o$ is the pressure ratio and $\kappa_o=4(2-\alpha)/\alpha \text{Kn}_o$ with $\text{Kn}_o=\mu_0/(Rp_o)\sqrt{\pi R_sT/2}$ being the Knudsen number at the outlet of the microchannel and $\alpha$ the accommodation factor. We have chosen $\alpha=0.9$ in agreement with the values reported in the literature that spread from 0.86 to 0.93 for nitrogen-silica interface~\cite{lei2014revisiting}. Finally, it should be mentioned that, in contrast to the DSMC data, the analytical formula (Eq.~\ref{eq:pressure}) completely ignores the pressure drop phenomenon at the fiber inlet. 


\section*{Local velocity gas flow distribution}

To perform local flow velocity sensing, the trapping lasers are rapidly switched on and off. In absence of the trapping lasers, the particle experiences a drag force $F_d=-\gamma v(z,r)$ where $v(z,r)$ is the local flow velocity profile. The particle accelerates towards the low pressure side until it reaches the gas velocity $v_\infty$. When the trapping laser is switched back on, the optical force overcomes the drag force and the particle is trapped again at an intensity maximum of the standing wave. Measuring the asymptotic particle velocity $v_\infty$ near position $z$ inside the HC-PCF is therefore a direct measure of the local gas velocity.

We perform the velocity flow measurement in the following way (see Fig.~\ref{fig:setup}C): at $t_0$, the particle is trapped in the 
optical trap at a position $z$. The trap is switched off during a time $\Delta t$ and the particle is dragged towards the vacuum chamber at pressure $p_o$ (Fig.~\ref{fig:setup}A). Typical values for $\Delta t$ are on the order of few ms. Longer switching off times would result in particle loss inside the fiber. The trap is switched back on at time $t_1$ and the particle is re-captured by an intensity maximum of the optical standing wave. We then repeat this protocol $N$ times and measure the total displacement $\Delta z$ with a CCD camera before and after the switching protocol. After each switching event, the particle is recaptured in an intensity maximum of the optical standing wave trap. The asymptotic velocity is then obtained as $v_\infty=\Delta z/(N\Delta t)$. The particle is then moved along the fiber using the conveyor belt and the protocol is repeated at a different position $z$ along the fiber. 

Note that the particle is not always recaptured at the center of the fiber ($r=0$) when the trap is switched on. This is due to the presence of higher order modes inside the fiber that modulate the radial trapping position of the particle~\cite{Grass2016}. We have characterized this effect and have estimated that the average radial trapping position is displaced by an amount $r_0=0.3 R$ with respect to the fiber axis, see Supplementary Materials. This trapping offset has to be taken into account in the DSMC results or in the analytical formula (Eq.~\ref{eq:v}) when comparing with the experimental results for the flow velocity.    

 \begin{figure}[t]
    \centering
    \includegraphics[width=0.9\columnwidth]{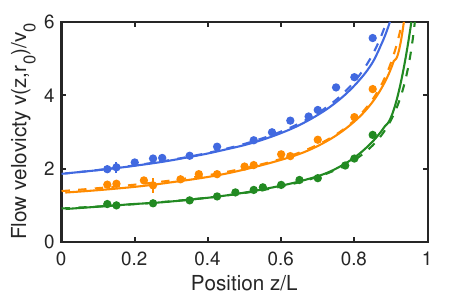}
    \caption{\textbf{Local flow velocity distribution.} Flow velocity profiles $v(z,r_0)/v_0$ measured with a levitated particle ($a=378$ nm) along the HC-PCF at the radial position $r_0=0.3R$. The velocity profiles are normalized to $v_0=v(0.15,r_0)$ at $p_i=300$ mbar with $v_0=0.032$ m/s. Inlet and outlet pressures and color coding are the same as in Fig.~\ref{fig:pressure}. Full circles are experimental data with error bars representing statistical uncertainties. Solid lines are DSMC  simulation results for flow velocity measured with an offset $r_0=0.3R$ from the fiber axis. Dashed lines are the analytical results from Eq.~\ref{eq:v}. Velocity flow data for different particle sizes can be found in the  Supplementary Materials (Fig. S8).}
    \label{fig:velocity}
\end{figure}

The gas velocity distributions are presented in Fig.~\ref{fig:velocity} for the same pressure settings used for the pressure distribution measurement. We notice that the gas flow accelerates as it moves down the HC-PCF. This is a direct consequence of the density drop caused by the decreasing pressure in the streamwise direction. In other words, the mean streamwise velocity must increase to maintain a constant mass flow. Our experimental results compare very well with the DSMC results calculated at a distance $r_0$ from the fiber axis. In addition, we show in Fig.~\ref{fig:velocity} the analytical expression for the flow velocity obtained for long microtubes and 
the slip flow boundary condition (see Supplementary Materials):
\begin{equation}
v(\tilde{z},\tilde{r})=-\frac{R^2}{4\mu_0}\frac{d\tilde{p}}{d\tilde{z}}\left(1-\tilde{r}^2+\kappa/2\right), \label{eq:v} 
\end{equation}
where $\tilde{p}$ is given by Eq.~\ref{eq:pressure}, $\tilde{r}=r/R$ and $\kappa=4(2-\alpha)/\alpha\text{Kn}$. 
\\
\section*{Discussion and outlook}
We have demonstrated a combined and reconfigurable pressure and velocity gas flow sensor using a levitated particle inside a hollow-core photonic crystal fiber. Our experimental and simulation data cover three flow regimes with Knudsen number spanning more than three orders of magnitude, from 0.02 to 67. In our results, we have probed the velocity flow distribution at a distance $r_0=0.3R$ from the center of the HC-PCF. The complete radial velocity profile can be mapped by adding a spatial light modulator to actively control the radial trapping position of the particle~\cite{Schmidt2013b}. 

It would be of interest to extend our results to temperature-driven gas flows when the HC-PCF is subjected to a temperature gradient. This leads to the counterintuitive phenomenon of thermal transpiration where the fluid creeps from colder to hotter regions~\cite{reynolds1879certain}. A direct application of thermal transpiration is the design of thermally driven micropumps (Knudsen pumps) operating without any moving mechanical part~\cite{lopez2019design}. Knudsen pumps are widely utilized for gas separators~\cite{nakaye2016demonstration}, gas analysis~\cite{ferran1996high} and micro combustors~\cite{yang2002combustion}.  

Our results open the door to new sensing applications in  microfluidics systems. Using liquid-filled hollow core photonic crystal fibers~\cite{garbos2011doppler}, our levitated sensors can probe in-situ and with high spatial resolution the pressure and velocity flow of gas or liquids in microtubes. This is particularly relevant for various microfluidic applications like heat transfer~\cite{kandlikar2005heat}, chemical reactions~\cite{tanimu2017heterogeneous} and bio-engineering with lab on a chip systems~\cite{haeberle2007microfluidic} where pressure-driven flows are the most commonly used delivery method. In these systems, accurate pressure control is crucial for loading and handling biological or chemical samples in microsystems~\cite{dittrich2006lab}.

\section*{Acknowledgments}
N.K. acknowledges support from the Austrian Science Fund (FWF): AY0095221, START.

\section*{Additional information}
The authors declare no competing financial interests.

\end{document}